# What do Indian Researchers download from Sci-Hub?

Vivek Kumar Singh[1], Satya Swarup Srichandan[1], Sujit Bhattacharya[2]

[1] Department of Computer Science, Banaras Hindu University, Varanasi-221005, India
[2] CSIR- National Institute of Science, Technology and Development Studies,
New Delhi-110012, India

**Abstract:** Recently three foreign academic publishers filed a case of copyright infringement against Sci-Hub and LibGen before the Delhi High Court and prayed for complete blocking these websites in India. In this context, this paper attempted to assess the impact that blocking of Sci-Hub may have on Indian research community. The download requests originating from India on a daily-basis are counted, geotagged and analysed by discipline, publisher, country and publication year etc. Results indicate that blocking Sci-Hub in India may actually hurt Indian research community in a significant way.

## Introduction

Recently three foreign academic publishers (Elsevier, Wiley and American Chemical Society) filed a case of copyright infringement against Sci-Hub and LibGen before the Delhi High Court and prayed for complete blocking these websites in India through a so-called dynamic injunction (Scaria, 2020). The matter is being heard by the Court and if the petition succeeds, these websites may face similar action to what happened in United States in 2017 (Schiermeier, 2017 a, 2017b). While many criticize Sci-Hub for copyright violation and threatening economic viability of publishers; a large number of people in academic and publishing community appreciate Sci-Hub for providing access to knowledge generated by the scientific community (Travis, 2016). Perhaps this may be the reason that many knowledge societies and non-government organizations in India are opposed to blocking these websites.

Sci-Hub was founded by Alexandra Elbakyan in 2011 in Kazakhstan in response to the high cost of access to research papers that are behind paywalls. It is considered as a pirate site that that provides free access to millions of research papers, without regard to copyright. Alexndra Elbakyan, the founder of Sci-Hub calls it as a "*true solution to open access*" Elbakyan (2016). It is believed that Sci-Hub contains more than 76 million academic articles (Himmelstein et al., 2017). Bohannon (2016) worked with Alexandara Elbakyan, to obtain the access log of Sci-Hub and analysed it. He observed that Sci-Hub download activity is spread across the world, including developed, developing and under-developed countries. In fact, in response to his question "*Who's downloading pirated papers from Sci-Hub*", he responds "*Everyone*".

Many argue that the outcome of the case against Sci-Hub and LibGen may have long-term consequence to research and education in India (Trivedi, 2021) and that blocking Sci-Hub may actually hurt national interest (Pai, 2020). However, there is no existing quantitative analysis on what number of research papers are actually downloaded by Indian researchers from Sci-Hub. It is in this context that we tried to find out how many research papers do Indian researchers download from Sci-Hub and from which places these download requests originate. The distribution of research papers downloaded among different publishers, publication year, discipline etc. are also analysed. The analytical results present very useful and interesting insight on the usage of Sci-Hub by Indian researchers, which in turn helps in assessment of the impact that blocking of Sci-Hub may have on Indian research community.

**Data and Method**

The main data for analysis is the Sci-Hub access log for the year 2017 (Tzovaras, 2018). This access log provides details of download requests received by the Sci-Hub website during the year 2017. The access log has a total of 150,875,861 entries, indicating that during the year 2017, more than 150 million download requests have been served by Sci-Hub. The access log has entries for a total of 329 days during 01.01.2017 to 31.12.2017, with log entries missing for 36 days. The access log has following fields:

```
Timestamp (yyyy-mm-dd-hh:mm:ss), DOI, IP Identifier, User
identifier, Country according to GeoIP, City according to
GeoIP, Latitude, Longitude.
```

The access log has been mined to find out download entries pertaining to India. For this purpose, all those entries that had 'Country according to GeoIP' as India were identified. It was found that out of 150,875,861 log entries, a total of 13,144,241 log entries corresponds to download requests from India, i.e., 8.7% of the download requests from Sci-Hub during the year 2017 originated from India. The 'Timestamp' field in access log was analysed to find out per day download activity from India. The 'latitude' and 'longitude' information were then used to geotag the download requests on the Indian geographical map. Finally, the 'DOI' information was used to get additional data (such as publisher, author affiliation, year of publication, field of research, open access status etc.) from Dimensions database for all the research papers downloaded. This information was then used to analyse the download activity by publisher, discipline, journal, open access status etc.

**How many research papers do Indian researchers download from Sci-Hub?**

The analysis of access log shows that a total of 13,144,241 download requests (8.7% of the total log entries) originated from India. These 13,144,241 download requests were for a total of 5,797,188 unique research papers. Thus, Indian researchers used Sci-Hub to download more than 5 million unique research papers amounting to a total of more than 13 million downloads. We have analysed and plotted the download activity from India on a daily-basis in **Figure 1**. Here the x-axis represents days and the y-axis represents the number of downloads. It can be seen that an average of 39,952 download requests are served by Sci-Hub on a daily-basis.

As a next step, we tried to measure the download activity on weekdays (Monday to Friday) as well as on weekends (Saturday & Sunday). **Figure 2** shows the average downloads on different days of the week. It can be seen that the average download requests on weekdays are much higher than that on weekends. This is perhaps an indication that Indian researchers are not only using Sci-Hub to download research papers on weekends (when the campuses are generally closed), but also a major part of the download is happening on the working days in the week. This implies that Sci-Hub is not only accessed by researchers from their homes but also from their workplaces.

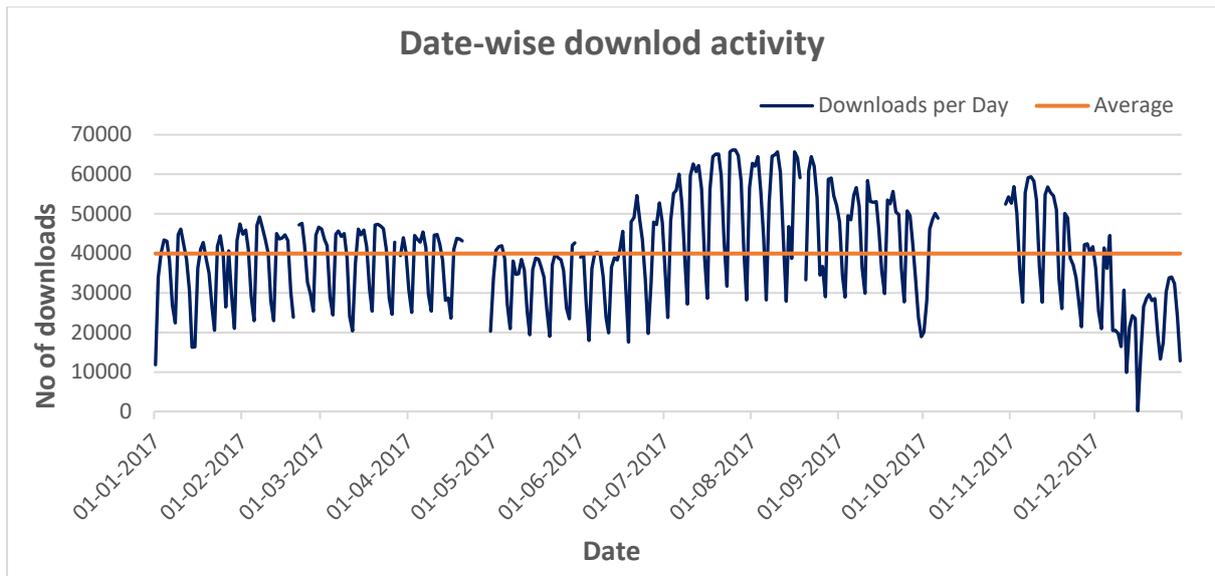

Figure 1: Date-wise number of downloads in the year 2017

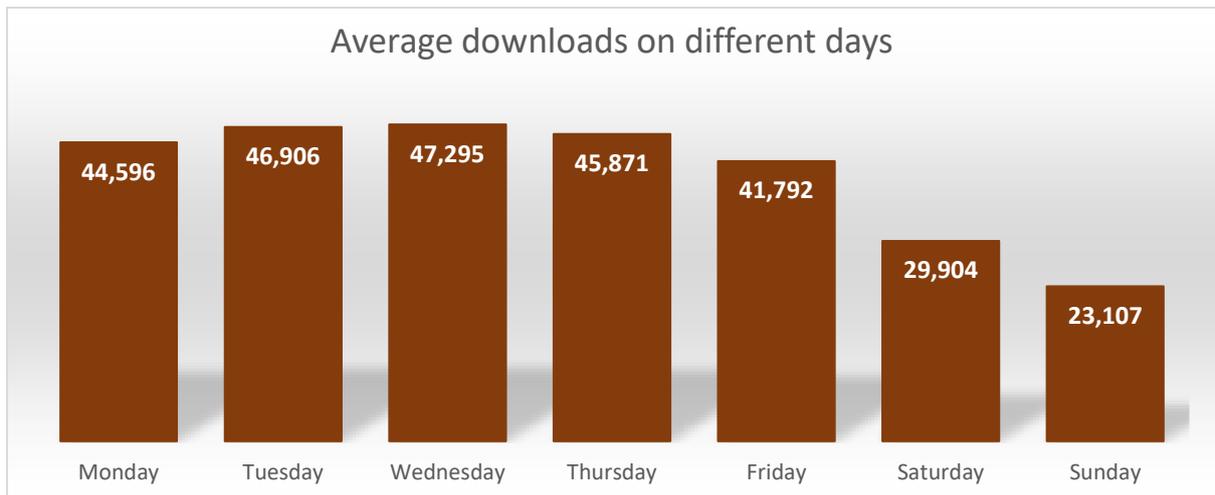

Figure 2: Average downloads for different days of a week during 2017

**Where in India do the downloads come from?**

The 'latitude' and 'longitude' information from the access log has been extracted and the corresponding places are geotagged on Indian geographical map. **Figure 3** shows the geotagged map of download activity originating from India. We can see that the download requests are distributed across different prats of India. A large number of requests originate from the major urban areas of New Delhi, Chennai, Bengaluru, Hyderabad, Mumbai, Gurugram, Pune, Kolkata, Ahmedabad etc. In terms of regions, it can be seen that there is high download activity originating from Delhi-Punjab-Haryana region, east coast regions of Maharashtra-Kerala, and west coast regions of Tamil Nadu- Andhra Pradesh- West Bengal. States of Gujrat, Uttar Pradesh and Bihar also have significant download requests. Thus, in general the download requests to Sci-Hub originate from almost all major populous parts of India.

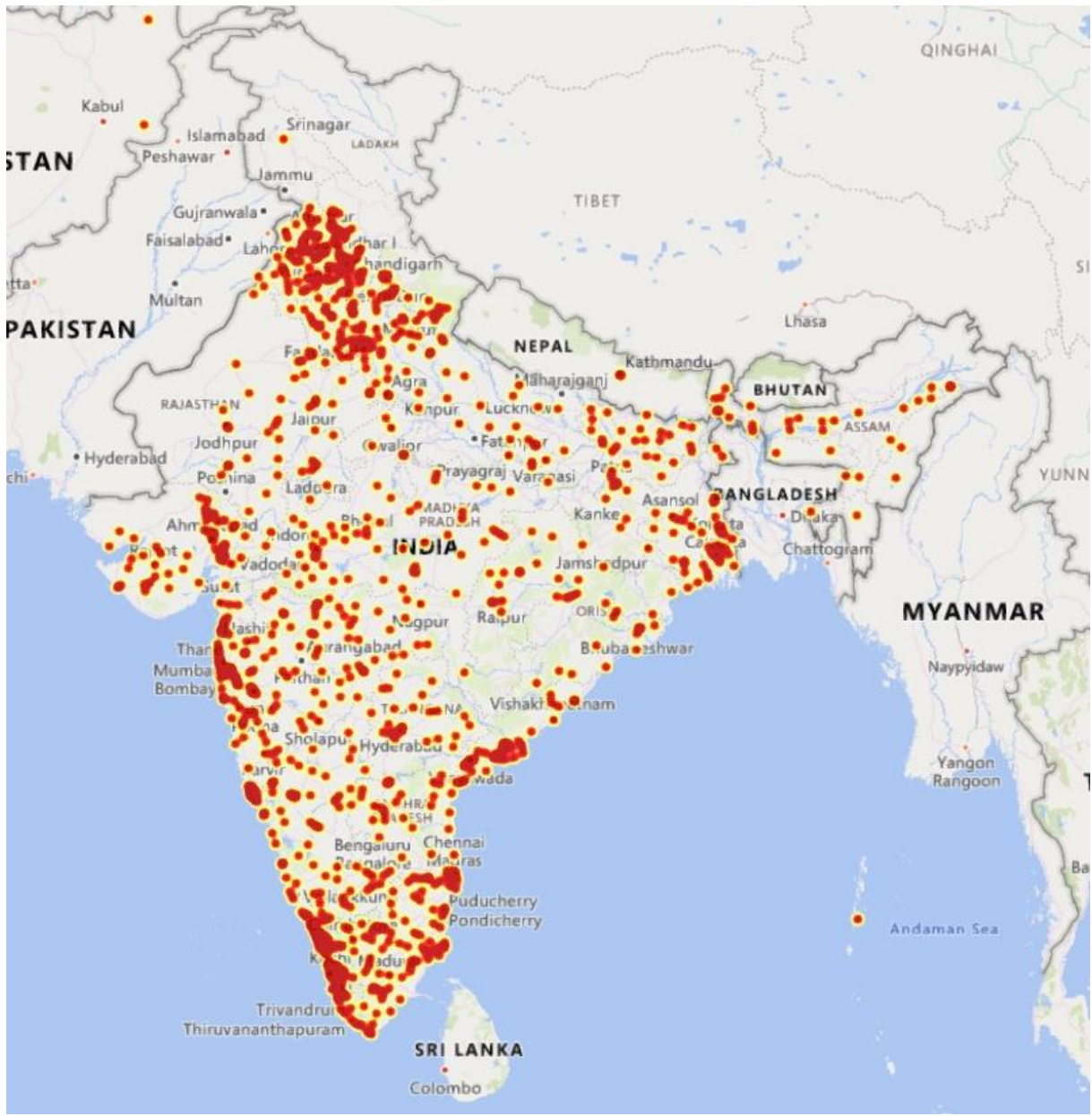

Figure 3: Geotagging the origin of download requests

**Which research papers are Indian researchers downloading?**

In order to further analyse what research papers, do Indian researchers download from Sci-Hub, we have obtained additional publication metadata from the Dimensions database for the DOIs occurring in access log. The Dimensions database has been used, owing to its large coverage of journals, among the popular databases (see Singh et al., 2021). The 13,144,241 access log entries correspond to a total of 5,797,188 unique DOIs. For each of these DOIs, Dimensions database was queried to obtain publication metadata. The publication metadata for a total of 5,688,915 DOIs was obtained from Dimensions database. The obtained metadata included details of journal, publisher, author affiliation country, publication year, fields of research, open access status etc. This metadata was analysed to understand distribution of the downloaded research papers by publication year, discipline, publisher etc. The major countries, journals and open access status of downloaded research papers are also identified.

*First of all*, the publication distribution of research papers was obtained. **Figure 4** shows the publication year distribution of the research papers downloaded. It can be seen that a major part of the research papers downloaded are for recent period (2000-2016). The download count contains more than 100,000 papers for all the years from 2000 to 2016. Some papers with publication year after 2016 are also there in the download log. There are, however, other research papers published before 2000 that are also downloaded from Sci-Hub. Out of the total research papers downloaded, 4,010,781 papers (70.5% of the total) are for the period 2000-16 and 1,335,885 papers (23.5% of the total) are for publication years before 2000. Sci-Hub is thus used to download not only the recently published research papers but also other research papers published before, though relatively recent papers are downloaded more.

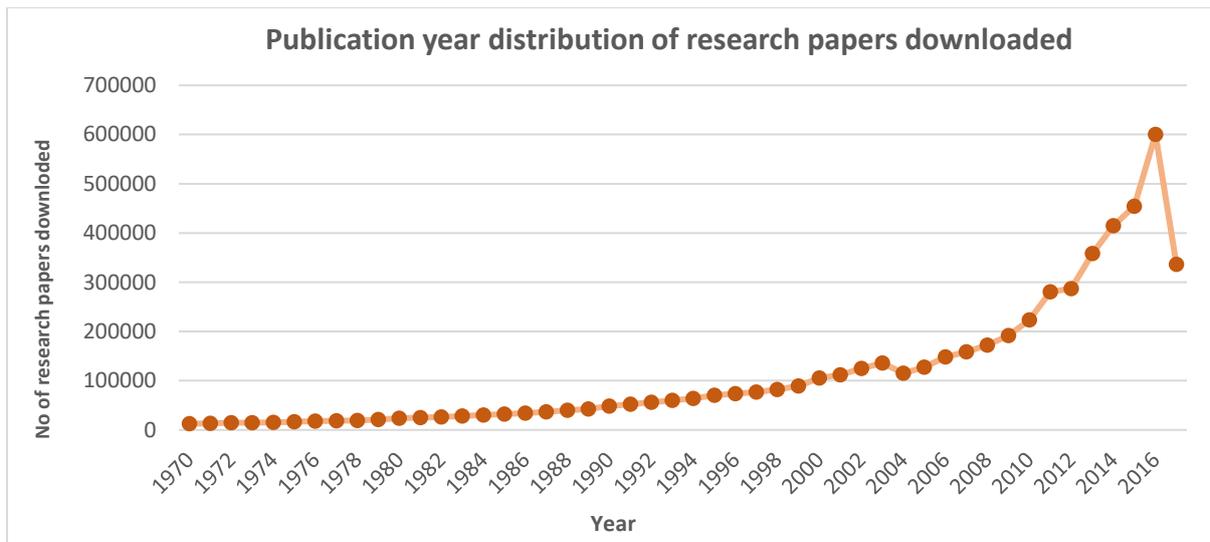

Figure 4: Publication-year distribution of research papers downloaded

*Secondly*, the disciplinary distribution of research papers downloaded was obtained. **Figure 5** shows the proportionate distribution of research papers downloaded in the 22 major subject areas of Dimensions database. It can be observed that Engineering (21%), Medical & Health Sciences (19%), Chemical Sciences (15%), and Information & Computing Sciences (11%) account for the major share of research papers downloaded. These four disciplines taken together account for 66% of the total research papers downloaded. Other major disciplines include Biological Sciences (8%), Physical Sciences (5%) and Technology (5%). Thus, majority of the research papers downloaded are from Engineering and different sciences.

*Thirdly*, the publisher-wise distribution of research papers downloaded was obtained. **Figure 6** shows the proportionate share of research papers from different publishers downloaded from Sci-Hub. It can be seen that IEEE (22%), Springer Nature (14%), Elsevier (12%) and Wiley (12%) are the major publishers, whose articles are downloaded from Sci-Hub. These four major publishers taken together account for 60% of the total research papers downloaded. Among these publishers, Elsevier, Wiley and American Chemical Society are the three publishers who have filed the lawsuit against Sci-Hub in India. These three publishers together account for 29% of the total research papers downloaded.

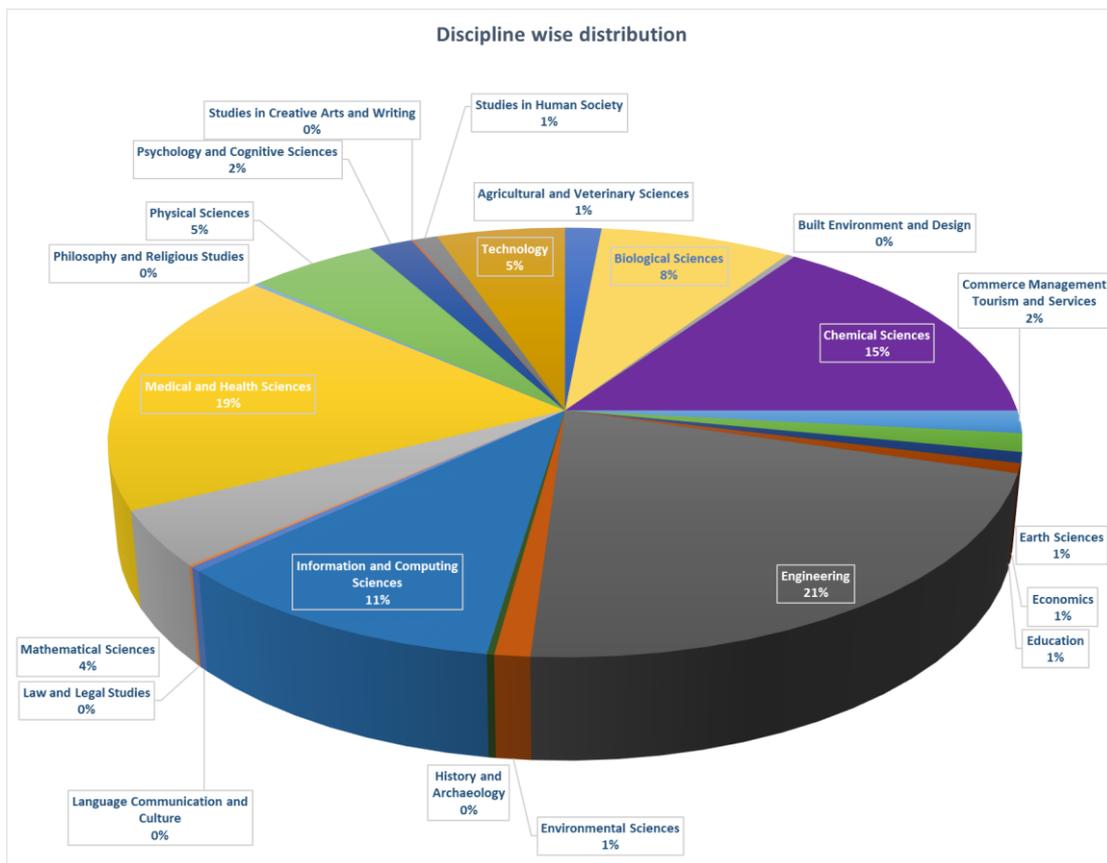

Figure 5: Discipline-wise distribution of research papers downloaded

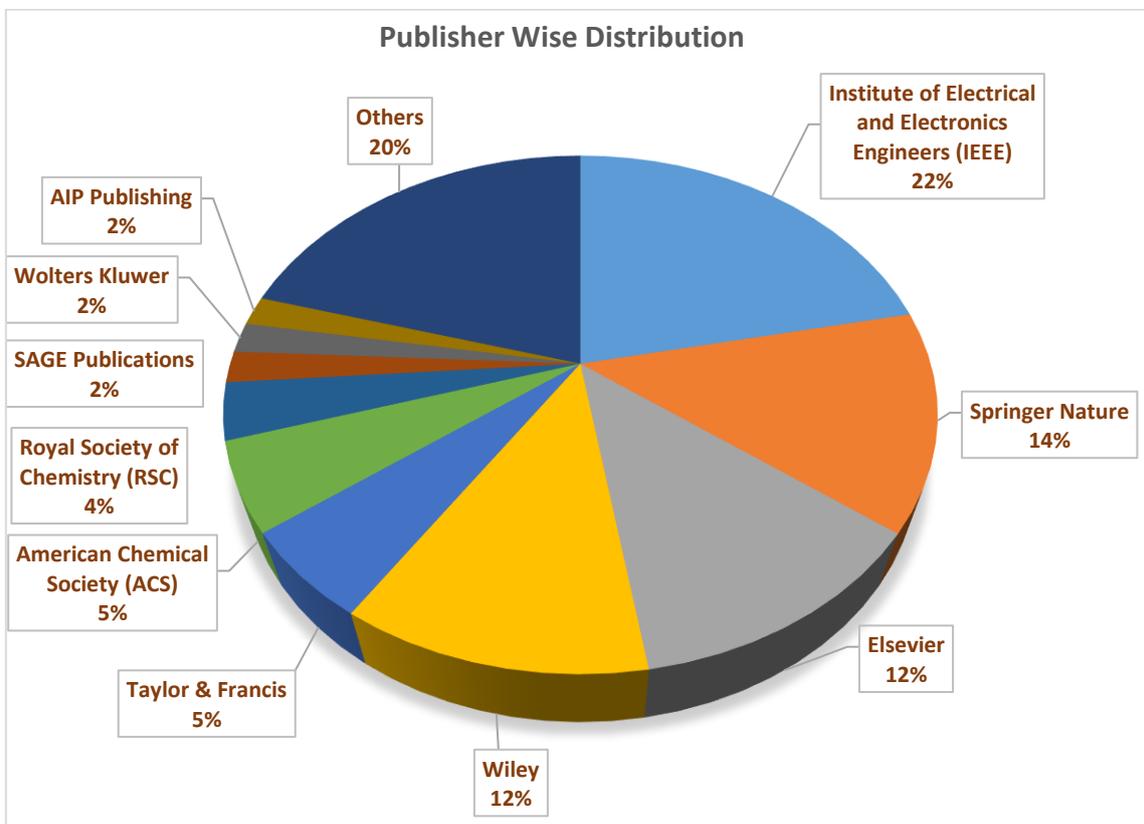

Figure 6: Publisher-wise distribution of research papers downloaded

*Fourthly*, the country-wise distribution of research papers downloaded was obtained. **Table 1** shows the list of top 25 countries with major share of research papers downloaded. These include United States (21.92%), China (9.75%), India (6.99%), United Kingdom (5.24%), Japan (3.96%), Germany (3.90%) etc. Incidentally, these are also the major countries producing higher research output globally. We see that Indian researchers used Sci-Hub to download 397,691 research papers published by Indian authors.

Table 1: Country-wise distribution of research papers downloaded (top 25)

| Country | No of research papers | Percentage |
| --- | --- | --- |
| United States | 1,247,260 | 21.92 % |
| China | 555,067 | 9.75 % |
| India | 397,691 | 6.99 % |
| United Kingdom | 298,218 | 5.24 % |
| Japan | 225,295 | 3.96 % |
| Germany | 222,311 | 3.90 % |
| Canada | 156,625 | 2.75 % |
| France | 141,912 | 2.49 % |
| Italy | 128,421 | 2.25 % |
| South Korea | 114,368 | 2.01 % |
| Australia | 112,547 | 1.97 % |
| Spain | 90,080 | 1.58 % |
| Taiwan | 80,355 | 1.41 % |
| Netherlands | 72,544 | 1.27 % |
| Brazil | 59,433 | 1.04 % |
| Iran | 58,289 | 1.02 % |
| Switzerland | 53,822 | 0.94 % |
| Sweden | 52,167 | 0.91 % |
| Turkey | 51,297 | 0.90 % |
| Russia | 44,167 | 0.77 % |
| Singapore | 38,689 | 0.68 % |
| Belgium | 36,379 | 0.63 % |
| Malaysia | 34,646 | 0.60 % |
| Poland | 32,938 | 0.57 % |
| Israel | 32,433 | 0.57 % |

*Fifthly*, the major journals for which a good number of research papers are downloaded from Sci-Hub are identified. Table 2 shows the top 25 journals arranged in descending order of research papers downloaded from them. We see that Journal of the American Chemical Society, RSC Advances and Proceedings of SPIE are the top three in the list. Multidisciplinary journals like Nature, Science and Scientific Reports are also included in the list.

Table 2: Top 25 Journals with number of research papers downloaded

| Journal | No of research papers | Percentage |
| --- | --- | --- |
| Journal of the American Chemical Society | 40,403 | 0.71 % |
| RSC Advances | 38,879 | 0.68 % |
| Proceedings of SPIE | 29,921 | 0.52 % |
| The Journal of Organic Chemistry | 26,058 | 0.45 % |
| Physical Review B | 25,614 | 0.45 % |
| Nature | 24,288 | 0.42 % |
| Journal of Applied Physics | 24,029 | 0.42 % |
| Applied Physics Letters | 22,783 | 0.40 % |
| Chemical Communications | 22,540 | 0.39 % |
| AIP Conference Proceedings | 20,792 | 0.36 % |
| Science | 20,137 | 0.35 % |
| Angewandte Chemie International Edition | 17,617 | 0.30 % |
| The Journal of Chemical Physics | 14,907 | 0.26 % |
| Journal of Medicinal Chemistry | 13,828 | 0.24 % |
| Journal of Agricultural and Food Chemistry | 13,735 | 0.24 % |
| Journal of Applied Polymer Science | 13,554 | 0.23 % |
| Analytical Chemistry | 13,522 | 0.23 % |
| The Journal of Physical Chemistry C | 13,180 | 0.23 % |
| Organic Letters | 13,003 | 0.22 % |
| Tetrahedron Letters | 11,914 | 0.20 % |
| The Lancet | 11,911 | 0.20 % |
| ACS Applied Materials & Interfaces | 11,750 | 0.20 % |
| Annual International Conference of the IEEE Engineering in Medicine and Biology Society (EMBC) | 11,724 | 0.20 % |
| Scientific Reports | 11,510 | 0.20 % |

*Finally*, we also tried to find out that what amount of research papers downloaded from Sci-Hub are also available in some form of open access. Table 3 shows that 18.46% of the total research papers downloaded from Sci-Hub are actually available in open access forms and are not locked in access behind a paywall. Out of the openly accessible research papers, 9.72% are in green open access (i.e., deposited in institutional or disciplinary repositories) and 4.94% and 2.43% are bronze and gold open access, respectively. Thus, we can see that Sci-Hub has been also used to download research papers that are otherwise available in open access forms.

Table 3: Openly accessible research papers downloaded from Sci-Hub

| Open Access Type | No of research papers | Percentage |
|---|---|---|
| green | 553,166 | 9.72 % |
| bronze | 281,429 | 4.94 % |
| gold | 138,314 | 2.43 % |
| hybrid | 77,711 | 1.36 % |
| **Total** | **1,050,620** | **18.46 %** |

**Discussion**

This article analysed the Sci-Hub access log to find out how many research papers are downloaded by Indian researchers from Sci-Hub, and where do these download requests originate from. It is found that Indian researchers have made more than 13 million downloads during the year 2017, averaging to a daily download of about 39,952 research papers. The average downloads are relatively higher on week days as compared to weekends. Further, the download requests are found to be distributed across different parts of India. The article also analysed the distribution of downloaded research papers in different publication years, disciplines, publishers etc. It is seen that a large proportion of papers downloaded are less than 20 years old. Further, research papers from different publishers and countries are downloaded through Sci-Hub. A good number of 397,691 Indian research papers are also downloaded by Indian researchers from Sci-Hub. About 18.46% of research papers downloaded from Sci-Hub are also available in different open access forms.

The recent lawsuit filed by the three foreign publishers against Sci-Hub and LibGen has attracted a lot of attention of Indian research community. Some intervention applications have also been filed by different knowledge societies and non-government organizations in the court in this matter. There are different kinds of arguments put forward in these applications as well as in other published literature on this matter. The foremost among these is that scientific papers are intellectual products of authors and institutions, and that publishers add no significant value. Publishers, on the contrary, continue to charge high profits with exorbitant subscription and APC charges. It is said that Sci-Hub and LibGen have emerged as counter-movements against the propertisation of scientific communication, and that they are in a way addressing the problem of inequalities in access. Therefore, blocking Sci-Hub and LibGen may actually disserve the public interest. Moreover, Sci-Hub and LibGen do not profit in any way from the access they provide. Some people argue that knowledge is a non-zero-sum game and hence access provided to research papers by Sci-Hub actually does not take away anything from the authors or institutions who publish the paper. There are also arguments that not all material on Sci-Hub and LibGen are copyrighted, and that blocking of these websites is practically a costly affair, involving public costs.

While, most of the arguments above are quite convincing and highlight very important aspects of the whole problem of access to scientific research, some take the position that blocking Sci-Hub may actually hurt national interest. One may, therefore, try to assess the impact that blocking Sci-Hub may have on Indian research ecosystem by looking at the analytical results. The analytical results show that 39,952 research papers are downloaded from Sci-Hub each day, which is a significant amount. Therefore, if Sci-Hub gets blocked, Indian researchers would not be able to download these papers. Even if we assume that 18.46% of these papers can be accessed through other open access rates, a significant number of 32,576 papers will still become inaccessible. Further, we see that Sci-Hub usage is distributed across different parts of the India, suggesting that the access inequalities exist in almost all parts of India. Therefore, blocking of Sci-Hub will impact researchers in the entire country. If a large number of researchers become unable to access scientific literature, it is bound to impact their research and productivity. It may be noted that the present analytical data is for the year 2017 and it is quite likely that the current usage of Sci-Hub will be much more than this. Therefore, it appears that blocking Sci-Hub may actually have long term consequences to research in India, as already suggested in some arguments. The entire situation and the current lawsuit are also an indication that India should now take a proactive step in forming broader coalition for negotiations with publishers for access to journals as well as on APC charges. At the same time, since we see that a good number of 397,691 Indian research papers are also accessed through Sci-Hub, therefore, it is necessary to strengthen mechanisms of institutional repositories (as also pointed out in Singh et al., (2020)) so that at least all research output from India can be accessed freely by Indian researchers.